\documentclass[12pt,preprint]{aastex}

\shorttitle{Predicting the Solar Cycle}
\shortauthors{Bushby \& Tobias}

\newcommand{\etal}{{\it et al }}

\begin{document}

\title{On Predicting the Solar Cycle using Mean-Field Models}

\author{Paul J. Bushby\altaffilmark{1}}
\email{P.J.Bushby@damtp.cam.ac.uk}

\author{Steven M. Tobias\altaffilmark{2}}
\email{smt@maths.leeds.ac.uk}

\altaffiltext{1}{DAMTP, Centre for Mathematical Sciences, University of
  Cambridge, Wilberforce Road, Cambridge CB3 0WA, U.K.}

\altaffiltext{2}{Department of Applied Mathematics, University of Leeds,
  Leeds LS2 9JT, U.K.}

\begin{abstract}
We discuss the difficulties of predicting the solar cycle using
mean-field models. Here we argue that these difficulties arise owing
to the significant modulation of the solar activity cycle, and that
this modulation arises owing to either stochastic or deterministic
processes. We analyse the implications for predictability in both of
these situations by considering two separate solar dynamo models. The
first model represents a stochastically-perturbed flux transport
dynamo. Here even very weak stochastic perturbations can give rise to
significant modulation in the activity cycle. This modulation leads to
a loss of predictability. In the second model, we neglect stochastic
effects and assume that generation of magnetic field in the Sun can be
described by a fully deterministic nonlinear mean-field model --- this
is a best case scenario for prediction. We designate the output from
this deterministic model (with parameters chosen to produce
chaotically modulated cycles) as a target timeseries that subsequent
deterministic mean-field models are required to predict. Long-term
prediction is impossible  even if a model that is correct in all
details is utilised in the prediction. Furthermore, we show that even
short-term prediction is impossible if there is a small discrepancy in
the input parameters from the fiducial model. This is the case even if
the predicting model has been tuned to reproduce the output of
previous cycles. Given the inherent uncertainties in determining  the
transport coefficients and nonlinear responses for mean-field models,
we argue that this makes predicting the solar cycle using the output
from such models impossible.
\end{abstract}


\section{Introduction}

Magnetic activity in the Sun is known to play a central role in
driving both long-term and short-term dynamics (Tobias 2002; Weiss
2002). The magnetic field is responsible for spectacular
events such as sunspots, solar flares, and coronal mass ejections, and for
heating the solar corona to high temperatures. Large-scale magnetic
activity is known to be dominated by the eleven year activity
cycle. This cycle has been systematically observed since the early
seventeenth century and its properties are well documented (see
e.g. Ossendrijver 2003). Of particular current interest is the impact
of magnetic activity on solar irradiance that might have significant
implications for the terrestrial climate (see Solanki \etal 2004).

Given the importance of solar activity, it is not surprising that
there has been a continued interest in understanding the mechanisms
responsible for generating the solar magnetic field. The Sun's
magnetic field is believed to be generated by a hydromagnetic dynamo
in which motion of the solar plasma (advection) is able to sustain a
magnetic field against the continued action of ohmic dissipation (see
e.g Moffatt 1978; Charbonneau 2005). Progress in understanding this
fundamental problem of solar magnetohydrodynamics is slow owing to the
difficulties of the dynamo problem. The extreme parameters of the
solar interior and the inherent three-dimensionality of the dynamo
problem make it impossible to solve the equations accurately on a
computer. Much effort has therefore focused on {\it mean-field} dynamo
models (Steenbeck, Krause \& R\"adler 1966; Krause \& R\"adler 1980),
which describe the evolution of the mean magnetic field,
parameterising the effects of the small-scale fields and flows in
terms of tensor transport coefficients. These transport coefficients
include $\alpha_{ij}$ (which leads to a regenerative term in the
mean-field equations --- the so-called $\alpha$-effect) and the
turbulent diffusivity ($\beta_{ijk}$). We stress here that there is no
mechanism within the theory for determining the form of these
coefficients, except for flows at low magnetic Reynolds number or with
short correlation time, and in solar models these are usually chosen
in a plausible but ad-hoc manner (often, for simplicity, adopting
isotropic representations in which $\alpha_{ij}=\alpha \delta_{ij}$
and $\beta_{ijk}=\beta\epsilon_{ijk}$). Much attention has been
focused upon determining these transport coefficients in both the
linear and nonlinear dynamo regimes from numerical simulations
(Cattaneo \& Hughes 1996; Brandenburg \& Subramanian 2005) but there
is still no consensus over the nature of these, even to within an order of
magnitude (see Courvoisier, Hughes \& Tobias 2006). Mean-field
models have, however, proved successful in providing illustrations of
the type of behaviour that might be expected to occur in the Sun (and
other stars). It is often argued that, {\it although these models have
no predictive power}, understanding the underlying mathematical form
of the equations can lead to the identification of robust patterns of
behaviour.

Many different models have been proposed for the solar dynamo. In the
distributed dynamo model, the $\alpha$-effect operates throughout the
convection zone and interacts with the latitudinal shear (or the
sub-surface shear layer, see Brandenburg 2005) to generate magnetic
field. Alternatively, the dynamo could be operating near the
tachocline, where an $\alpha$-effect might be driven either by a
tachocline-based instability or by turbulent convection. This, in
conjunction with the strong shear, could drive an ``interface''
dynamo (Parker 1993). Finally, there are flux transport models, in
which the (so-called) Babcock-Leighton mechanism produces an
$\alpha$-effect (or source term) at the surface. This surface
$\alpha$-effect is coupled to the radial shear in the tachocline
(where another $\alpha$-effect may be operating) via a meridional flow
(Choudhuri, Sch\"ussler \& Dikpati 1995; Dikpati \& Charbonneau
1999). The relative merits of these models are discussed elsewhere in
the literature (see, e.g. Charbonneau 2005) --- the only comment we
make here is that this plethora of models arises because of the lack
of available constraints on the form of the transport coefficients in
the mean-field formalism. We note further that it is not clear that
any of the above scenarios capture the essential dynamo processes
correctly or that these processes can ever be captured by a mean-field
model.

It is also possible to construct predictions of solar activity without
using dynamo theory, and there is a long literature describing these
predictive methods (see e.g. Zhang 1996; Hathaway, Wilson \& Reichmann
1999; Sello 2003; Zhao \etal 2004; Saba, Strong \& Slater 2005). One class
of prediction techniques uses statistical and timeseries analysis
methods (see e.g. Tong 1995 for details). These methods, which are
also applicable in many other areas of physics, vary in complexity
from simple linear methods to methods that use dynamical systems
theory to reconstruct nonlinear attractors in phase space. However,
these methods have the drawback that they do not utilise any of the
``physics'' of the problem. Predictions can also be made by using
precursor methods (see e.g. Schatten 2002), which do utilise some of
the physical features of the system in addition to the timeseries data.

In recent papers (Dikpati, de Toma \& Gilman 2006; Dikpati \& Gilman
2006), an attempt has been made to unify these two approaches by
utilising a mean-field model in order to make predictions about the
future activity of the Sun. These papers describe an axisymmetric,
mean-field model of a flux transport dynamo. Here the authors make use
of the observations of magnetic flux at the solar surface to feed into
a model of solar activity.  The flux that is observed at the solar
surface is advected by a parameterised meridional flow (which can be
observed down to a certain depth) and interacts with a differential
rotation profile that has been inferred from helioseismology. The
magnetic flux also interacts with turbulence, the effects of which are
parameterised by certain turbulent transport coefficients
(representing the turbulent diffusivity and the $\alpha$-effect). As
with all current mean field models these turbulent transport effects
have been parameterised in a plausible but ad-hoc manner, and are
unconstrained by observations and indeed theory. The simplest predictive scheme
proposed by Dikpati {\it et al} (2006) therefore takes the form of a
parameterised linear system forced by boundary observations. The
implicit underlying philosophy here is that by reducing the correct
physics for the generation of the solar activity cycle (i.e. a
nonlinear self-excited dynamo) to such a scheme, predictions about
future solar activity can be made. 

In this paper we shall investigate the predictability of various
dynamo models. We demonstrate that even when all the nonlinear
physics of the solar dynamo is removed, problems remain for prediction
owing to the increased importance of stochastic effects --- even very
weak stochastic perturbations can produce significant modulation in
these linear-type models. We also discuss the best-case
scenario for prediction where stochastic effects can be ignored, and
demonstrate that in these cases prediction is still difficult owing to
uncertainties in the input parameters of these parameterised
mean-field models.   

The paper is organised as follows. In the next section we describe (in
a general way) the importance of modulation and the role of
stochasticity and nonlinearity in solar dynamo models. In section 3 we
investigate a flux transport model and demonstrate how the presence of
even extremely weak noise can render predictions useless. In section 4
we consider the ``best-case" scenario for prediction where noise does
not play a role in the modulation --- we demonstrate that more
accurate prediction schemes may arise by using basic timeseries
analysis techniques rather than from constructing mean-field models of
the solar cycle. Finally, in section 5 we discuss the implications of
our work for predictions of the solar cycle.

\section{Problems for prediction and mechanisms for modulation}

In this section, we discuss the problems that must be overcome by
schemes designed to yield a prediction of future solar magnetic
activity. Some of these problems arise owing to the nature of solar
magnetic activity whilst others arise from the lack of a detailed
theory that is capable of describing solar magnetic activity in such
extreme conditions as those that exist in the solar interior.

It is clear that if the solar cycle were strictly periodic, with a
constant amplitude, then it would be straightforward to predict future
behaviour. However, all measurements of solar magnetic activity (both
direct observations and evidence from proxy data) indicate that the
variations in the magnetic activity do not follow a periodic
pattern. Departures from periodicity may be driven either by
perturbations or by modulation. For the case of a weakly perturbed
periodic system, the dynamics is essentially captured by the periodic
signal, with the small perturbations playing a secondary role. We
distinguish this behaviour from a modulated signal in which there are
significant departures from periodicity (often occurring on longer
timescales), with large variations in the observed amplitude of the
signal. All the evidence from direct observations indicates that the
solar cycle is strongly modulated. The amplitude of the solar cycle
varies enormously over long timescales, an extreme example of this
modulation was a period of severely reduced activity in the
seventeenth century known as the Maunder Minimum. Proxy data from
records of terrestrial isotopes, such as $^{10}$Be and $^{14}$C (see
e.g. Beer 2000, Weiss \& Tobias 2000, Wagner {\it et al} 2001),
demonstrate that this modulation has been a characteristic feature of
the solar magnetic activity over (at least) the last 20,000 years.

Mathematically there are only two possible sources for this strong
modulation of the basic solar cycle (Tobias 2002). The modulation may
arise either as a result of stochastic effects (see e.g.  Ossendrijver
\& Hoyng 1996) or by deterministic processes (see e.g. Tobias, Weiss
\& Kirk 1995). In this context we define deterministic processes to be
those that {\it are} captured by the differential equations of dynamo
theory, with no random elements. Stochastic processes are those
that occur on an unresolved length or timescale, and so can not be
described by the differential equations without including a random
element into the model.

It is well known that stochastic modulation can arise even if the
deterministic physics that leads to the production of the basic cycle
is essentially linear. This parameter regime is generally considered to be a
good one for prediction, since any nonlinear effects are only playing
a secondary role. However, in this stochastically-perturbed case, the
small random fluctuations that lead to the modulation will have large
short-term effects and render prediction extremely difficult, if not
impossible. Conversely, if the modulation arises purely as a result of
deterministic processes, then the underlying physics is nonlinear (or
potentially non-autonomous) and this leads to difficulty in prediction
owing to the possible presence of deterministic chaos and (more
importantly) the difficulty of constructing accurate nonlinear models
with large numbers of degrees of freedom. 

In the next two sections we demonstrate the problems for prediction
for dynamo models in both of the classes described above. In the next
section we describe a flux transport model of the same type as the one
used in the prediction scheme of Dikpati {\it et al} (2006) and we
demonstrate that even very small random fluctuations can produce
significant modulation, leading to extreme difficulties for
prediction. We then, in section 4, go on to describe a model where the
modulation arises owing to the presence of deterministic chaos and show that in
this case, prediction using model fitting is a poor way to proceed, but some
prediction is possible if it is possible to reconstruct the attractor
for activity. 

\section{Prediction using a stochastically-perturbed flux transport
  dynamo model} 

\subsection{The dynamo model}

We assume initially that the modulated solar magnetic activity can be
described by a stochastically-perturbed mean-field
dynamo model. In this model, nonlinear effects are playing a secondary
role, and all the modulation is being driven by the stochastic
effects. The aim of this section is to assess whether or not models
of this type can be used to make meaningful predictions of the solar
magnetic activity. In these models, the evolution of the large-scale
magnetic field is described by the standard mean-field equation (see,
for example, Moffatt 1978),   

\begin{equation}
\frac{\partial \mathbf{B}}{\partial t} = \nabla \times \left(
\alpha\mathbf{B}+\mathbf{U} \times \mathbf{B} - \beta \nabla \times \mathbf{B} \right).\label{eqn:1}
\end{equation}

\noindent Here, $\mathbf{B}$ represents the large-scale magnetic
field and $\mathbf{U}$ corresponds to the mean velocity field, $\beta$ is
the (turbulent) magnetic diffusivity, and the $\alpha\mathbf{B}$ term
corresponds to the mean-field $\alpha$-effect. Using the well-known
$\alpha\omega$ approximation, we solve this equation numerically in an
axisymmetric spherical shell ($0.6R_{\odot} \le r \le
R_{\odot}$ and $0 \le \theta \le \pi$). In solving
Equation~(\ref{eqn:1}) we need to ensure that $\mathbf{B}$ remains
solenoidal (i.e. $\nabla \cdot \mathbf{B}=0$). To achieve
this, we decompose the magnetic field into its poloidal and toroidal
components,  

\begin{equation}
\mathbf{B}= B(r,\theta,t)\mathbf{e_{\phi}} + \nabla \times \left(
A(r,\theta,t)\mathbf{e_{\phi}}\right),\label{eqn:2}
\end{equation}

\noindent where $B(r,\theta,t)$ denotes the toroidal (azimuthal) field
component and the scalar potential $A(r,\theta,t)$ relates to the
poloidal component of the magnetic field. So, rather than solving
Equation~(\ref{eqn:1}) directly, the problem has been reduced to solving two
coupled partial differential equations for the scalar quantities
$A(r,\theta,t)$ and $B(r,\theta,t)$. We adopt idealised boundary
conditions, in which $A=B=0$ at $\theta=0$ and $\theta=\pi$ and
$r=0.6R_{\odot}$ and $A$ and $B$ are smoothly matched to a potential
field at $r=R_{\odot}$.  

\par This particular dynamo model is closely related to the flux
transport model described by Dikpati \& Charbonneau (1999). The
large-scale velocity field, $\mathbf{U}$, is given by    

\begin{equation}
\mathbf{U}= u_r(r,\theta)\mathbf{e_{r}} +
u_{\theta}(r,\theta)\mathbf{e_{\theta}} +
\Omega(r,\theta)r\sin\theta\mathbf{e_{\phi}},\label{eqn:3}
\end{equation}

\noindent where $\Omega(r,\theta)$ is a prescribed analytic fit to the
helioseismologically-determined solar rotation profile (see, for
example, Bushby 2006) and $u_r$ and $u_{\theta}$ correspond to a
prescribed meridional circulation. We assume that the meridional
circulation pattern in each hemisphere consists of a single cell, with
a polewards flow at the surface and an (unobservable) equatorwards
flow at the base of the convection zone --- the flow is confined to the
region $R_b \le r \le R_{\odot}$. The functional form that we adopt
for this flow is similar in form to the one described by Dikpati \&
Charbonneau (1999),  

\begin{eqnarray}
u_r(r,\theta)&=&U_o \left( \frac{R_{\odot}}{r} \right)^2 \left[
  -\frac{2}{3} + \frac{1}{2}c_1\xi^{0.5}-
  \frac{4}{9}c_2\xi^{0.75}\right] \xi \sin \theta \left( 3\cos^2
\theta - \sin^2 \theta \right),\label{eqn:4}\\
u_{\theta}(r,\theta)&=&U_o \left( \frac{R_{\odot}}{r} \right)^3 \left[
  -1 + c_1\xi^{0.5}-c_2\xi^{0.75}\right] \cos \theta \sin^2
\theta,\label{eqn:5}  
\end{eqnarray}
 
\noindent where $\xi(r)=[(R_{\odot}/r)-1]$, $c_1=4[\xi(R_b)]^{-0.5}$,
$c_1=3[\xi(R_b)]^{-0.75}$, and $U_o$ is some characteristic flow
speed. This flow pattern can be stochastically perturbed by setting
$R_b=0.7R_{\odot}+\epsilon(t)$, where $\epsilon(t)$ is a time-dependent,
randomly fluctuating variable in the range $-0.005R_{\odot} \le
\epsilon \le 0.005R_{\odot}$. The aim here is to assess whether or not
such weak stochastic variations in the flow pattern could give rise to
significant modulation in the activity cycle, and if so what are the
consequences for prediction.

\par In order to complete the specification of the model, we need to
choose plausible functional forms for the $\alpha$-effect and the
turbulent magnetic diffusivity. It should be emphasised again that
these mean-field coefficients are poorly constrained by theory and
observations, although plausible assumptions can be made. Defining
$\beta_o$ to be a characteristic value of the turbulent magnetic
diffusivity within the solar convection zone, we adopt a similar
spherically-symmetric profile to that adopted by Dikpati \&
Charbonneau (1999),  

\begin{equation}
\beta(r) = \frac{1}{2} (\beta_o-\beta_c)\left[ 1 + \mbox{erf}\left(\frac{r-0.7R_{\odot}}{0.025R_{\odot}} \right)\right] +
  \beta_c, \label{eqn:6}
\end{equation}

\noindent where erf corresponds to the error function and $\beta_c$
(here taken to be $1\%$ of $\beta_o$) represents the magnetic
diffusivity below the turbulent convection zone. Following Dikpati
\& Charbonneau (1999), rather than prescribing a simple functional
form for $\alpha$ we neglect the $\alpha$-effect term in the toroidal
($B$) field equation and replace the corresponding $\alpha B$ term in
the poloidal ($A$) equation by a non-local, nonlinear source of
poloidal flux,  

\begin{eqnarray}
S(r,\theta,t)&=&\frac{S_o}{2}\left[ 1 +
  \mbox{erf}\left(\frac{r-0.95R_{\odot}}{0.01R_{\odot}} \right)\right]
\left[ 1 - \mbox{erf}\left(\frac{r-R_{\odot}}{0.01R_{\odot}}
  \right)\right] \\ \nonumber &&\left[
  1+\left(\frac{B(0.7R_{\odot},\theta,t)}{B_o}\right)^2\right]^{-1}\sin
  \theta \cos \theta B(0.7R_{\odot},\theta,t).\label{eqn:7}
\end{eqnarray}

\noindent Here, $S_o$ is a characteristic value of this poloidal
source and $B_o$ represents the (somewhat arbitrarily chosen) field
strength at which this non-local source becomes suppressed by the
magnetic field. This source term parameterises the contribution to the
poloidal magnetic flux due to the decay of active regions --- the
non-locality reflects the fact that active regions are believed to
form as the result of buoyant magnetic flux rising from the base of
the convection zone to the solar photosphere. See Dikpati \&
Charbonneau (1999) for a more detailed discussion of this source term, though
again it must be stressed that the functional form and the nonlinear dependence
are chosen in a plausible yet ad-hoc manner. 

\subsection{Numerical results}

In order to carry out numerical simulations, we first
non-dimensionalise this flux transport model. By using scalings similar
to those described by Dikpati \& Charbonneau (1999), it can be shown
that the model solutions are fully determined by two non-dimensional
parameters (once other parameters such as $B_o$ have been selected).
Denoting the equatorial angular velocity at the solar
surface by $\Omega_{eq}$, these non-dimensional parameters are the
Dynamo number, $D=S_o\Omega_{eq}R_{\odot}^3/\beta_o^2$, and the
magnetic Reynolds number corresponding to the meridional flow,
$Re=U_oR_{\odot}/\beta_o$. Here, we set $D=7 \times 10^6$ and
$Re=5600$. In the absence of stochastic noise, this set
of parameters produces a strong circulation-dominated dynamo in which the
magnetic energy is a periodic function of time. Although the dynamo
number is not weakly supercritical, nonlinear effects are not strong
enough here to produce a modulated activity cycle --- the primary role
of the nonlinearity is to prevent the unstable dynamo mode from
growing exponentially. We term such a model a ``linear-type'' model. 

\par When weak stochastic effects are included in the model, the
resulting activity cycle is indeed weakly modulated. This is
illustrated in Figure~1, which shows the time-dependence of this
solution. The time-series clearly illustrates that, although the
amplitude of the ``cycle minimum'' only appears to be weakly
time-dependent, there are significant variations in the peak amplitude
of the magnetic energy time-series. These variations are qualitatively
similar to those observed by Charbonneau \& Dikpati (2000), who
considered large amplitude random fluctuations in the flow pattern
within the solar convection zone --- the peak amplitude of these
fluctuations was comparable with the peak amplitude of the flow. In this
particular model, we have shown that even very weak stochastic
variations in the centre of mass of the flow pattern can still produce
significantly modulated behaviour. These stochastic effects are
expected to become increasingly significant for dynamo numbers
approaching critical. So, these models are obviously highly sensitive
to the addition of stochastic noise.   

\par In the absence of stochastic noise, the attractor (in phase
space) for this solution is two-dimensional, and the future behaviour
of the solution at any instant in time is entirely determined by the
current position of the system on the attractor. The same is not true
when this system is perturbed by stochastic effects, and it clearly
becomes much more difficult to predict the future behaviour of the
system. Since the attractor of this stochastically perturbed solution
cannot be unambiguously defined, another possible way of assessing the
``predictability'' of this solution is to look for a correlation
between successive cycle maxima. Defining $T_n$ to be the magnitude of
the $n^{th}$ cycle maximum, Figure~2 shows $T_{n+1}$ as a function of
$T_n$. It is clear from this scatter plot that there is no obvious
correlation between the amplitudes of successive cycle maxima in this
case. Since the modulation is being driven entirely by random
stochastic forcing, this result is not surprising. This lack of
correlation suggests that the behaviour of previous cycles cannot be
used to infer the magnitude of the following one. This implies that
even weak stochastic effects may seriously reduce the possibilities
for solar cycle prediction in this linear-type regime.

\section{Predictions using a deterministic dynamo model}

\subsection{The dynamo model}

In the previous section, we demonstrated that even very weak
stochastic perturbations to the meridional flow pattern can lead to a
loss of predictability in a linear-type flux transport dynamo
model. In that model, the modulation of the activity cycle was driven
entirely by stochastic effects. As discussed in Section~2, the only other
possible scenario is that the observed modulation is driven by
nonlinear effects. This scenario, where the observed modulation is
deterministic in origin, is the ``best-case'' scenario for prediction,
as in this case the entirely unpredictable stochastic elements may be
ignored. We stress again that, given that solar magnetic activity is
significantly modulated, either deterministic or stochastic modulation must be
considered in any realistic model (predictive or otherwise) of the
solar cycle. So, in this section, we completely neglect stochastic effects and
assume that the observed (chaotic) modulation in the solar magnetic
activity can be described by a fully deterministic model in which any activity
modulation (e.g. solar-like ``Grand minima'') is driven entirely by
nonlinear effects. The model that we use was described in detail in
two recent papers (Bushby 2005, 2006), so we only present a brief
description here. The exact details of the model are unimportant for
our main conclusions.   

\par Like the flux transport dynamo model from the previous section,
this model describes an axisymmetric, mean-field,
$\alpha\omega$-dynamo in a spherical shell. Unlike the previous model,
this model represents an ``interface-like'' dynamo that is operating
primarily in the region around the base of the solar convection
zone. It is worth mentioning again that (as discussed in
the introduction) there is still no general consensus regarding which of
these dynamo scenarios is more likely to be an accurate representation
of the solar dynamo. For this interface-like dynamo model, we neglect
meridional motions, since they are poorly determined near the base of the solar
convection zone. Like several earlier models (e.g. Tobias 1997;
Moss \& Brooke 2000; Covas \etal 2000), this dynamo model includes the
feedback (via the azimuthal component of the Lorentz force) of the
mean magnetic field upon the differential rotation (Malkus \& Proctor
1975). This nonlinear feedback is a crucial element of the model and, in the
absence of stochastic effects, is the sole driver of modulation in the
magnetic activity cycle. Denoting this magnetically-driven velocity
perturbation by $V(r,\theta,t)$, the large-scale velocity field is
given by

\begin{equation}
\mathbf{U} = \left[\Omega(r,\theta)r\sin\theta +
  V(r,\theta,t)\right]\mathbf{e_{\phi}},
\label{eqn:8}
\end{equation}

\noindent where (as in the previous model) $\Omega(r,\theta)$
represents an analytic fit to the solar differential
rotation. Whilst the evolution of the large-scale magnetic field is
again governed by Equation~(1), an additional evolution equation is
required for the velocity perturbation, $V$. This equation is given by 

\begin{eqnarray}
\frac{\partial V}{\partial t}&=& \frac{1}{\mu_o
\rho} \left[(\nabla \times \mathbf{B}) \times \mathbf{B} \right]\cdot
\mathbf{e_{\phi}} +
 \frac{1}{r^3}\frac{\partial}{\partial r}\left[\nu r^4
\frac{\partial}{\partial r} \left(\frac{V}{r}\right)\right] \\
\nonumber &&+
\frac{1}{r^2 \sin^2 \theta}\frac{\partial}{\partial \theta}\left[\nu
  \sin^3 \theta \frac{\partial}{\partial \theta} \left(\frac{V}{\sin
\theta}\right)\right],\label{eqn:9}
\end{eqnarray}
 
\noindent where $\rho$ represents the fluid density (here taken to be
constant), $\mu_o$ is the permeability of free space and $\nu$
represents the (turbulent) fluid viscosity.

\par In order to complete the model, the spatial dependence of the
transport coefficients ($\alpha$, $\beta$ and $\nu$) must also be
specified. Again, we emphasise that there are no direct
observational constraints relating to these coefficients --- as noted
in the introduction, there is no consensus as to their form and there
is still a debate as to their order of magnitude (and even their
sign). Having said that, it is possible to make some plausible
assumptions for an ``interface-like'' dynamo model (see Bushby 2006
for more details). The precise choices of these parameters are unimportant for
our main conclusions.  

\par Having set up this model, it is possible to choose a set of
parameters so that the solutions {\it do} reproduce some salient
features of the solar dynamo (Bushby 2005, 2006). We stress here that,
although the parameters have been chosen in a plausible manner, this
dynamo model should not be regarded as an accurate representation of
the solar interior and is subject to many uncertainties. Furthermore
we stress again that this is the case with {\it all} mean-field solar
dynamo models. However we use this model as a useful tool to analyse
the possibility of producing predictive models of the solar cycle. We
proceed by choosing fiducial parameters and profiles for the turbulent
transport coefficients that lead to ``solar-type" magnetic activity,
with chaotically modulated cycles and recurrent ``Grand Minima''. We
then integrate this model forward in time to produce a timeseries and
designate this timeseries as the ``target" run, which any subsequent
model should be able to predict. This target run is shown in Figure~3,
which shows a timeseries of the activity together with a
reconstruction of the dynamo attractor in phase space. Although this
solution is chaotically modulated, it is certainly no more chaotic than
the equivalent attractor for the $^{10}$Be data, which is a well-known
proxy for solar magnetic activity (e.g. Beer 2000). Whilst the
nonlinear effects are significant enough to drive the modulation, they
are actually very difficult to detect. In this model, the cyclic
component of the fluctuations in the differential rotation (which are
driven by the nonlinear Lorentz force) are small compared with the
mean differential rotation. This is consistent with observations of
the (so called) torsional oscillations in the solar convection zone. Finally,
note once more that, since the modulation is driven entirely by
nonlinear effects, this model is specified exactly. 

\subsection{Numerical results}

The question is then posed as to whether {\it any} mean-field model
can be constructed that leads to meaningful predictions of the future
behaviour of the target run. Clearly the best chance for a mean-field
model being capable of predicting the future behaviour of the target
run is to use {\it the exact model} that led to the target run
data. Hence we test this model first, as all subsequent models will be
inferior to this. We proceed by setting the model parameters to be
those that generated the long test run, and consider the behaviour of
solutions that are started from very similar points on the
attractor. Some of the solutions are shown in Figure~4. This figure
shows clearly that although the predictor solutions are able to track
the target solution for a couple of activity cycles, the nature of the
solutions means that the predictors and target solution diverge
quickly after this time. This is not surprising behaviour. It is
well-known that chaotic solutions have a sensitive dependence on
initial conditions and that long-term prediction of such solutions is
fraught with problems (see e.g. Tong 1995). What is clear is that
simply using a model that is based upon mean-field theory will not
work in the long term {\it even if the model is correct in every
detail}. One might be able to predict one or two cycles ahead {\it if
one has solved the problem of constructing an exact representation of
the solar dynamo} but as noted above this is not an easy task.

We now turn to the related problem of short-term prediction. As
discussed in the introduction there are large uncertainties in the
form and amplitude of the input parameters for all mean-field models.
What we investigate here is whether these uncertainties lead to
significant difficulties in prediction even in the short term. Again
we examine the best case scenario and consider a mean-field model for
the predictor runs that has correctly parameterised the {\it form} of
all the input variables (differential rotation, $\alpha$-effect,
turbulent diffusivity and nonlinear response). In addition these
predictors have been given the correct input {\it values} for
all-but-one of the parameters. Hence the predictor models are exactly
the same as the target model with the exception of one input parameter
that has been altered by $5\%$. This would be a staggeringly good
representation should it be possible to achieve this for solar
activity. Furthermore we increase the chances of the predictor being
able to predict the future behaviour of the target solution by
matching the two timeseries over a number of cycles. This is analogous
to the procedure employed by Dikpati \etal (2006) who cite support for
their forecasting model by assuring that their model agrees with the
solar cycle data for eight solar cycles --- in reality this is not
difficult to achieve with enough model parameters at one's
disposal. Figure~5 shows the results of integrating the predictor
models for two different choices of incorrect parameter. Note that
even though the predictor has been designed to reproduce the target
over a number of cycles and that the predictor is very closely related
to the target, there is still a good chance that it can get the next
cycle incorrect, with significant errors in (particularly) the cycle
amplitude. There are also clear variations in the cycle period, which
obviously implies that the exact time between successive cycle maxima
is also an unpredictable feature of the system.   

We stress again that any mean-field model of solar activity includes
transport coefficients that are still uncertain possibly to an order of
magnitude (and certainly not to $5\%$ accuracy). Although the
incredible success of global and local helioseismology is placing
restrictions on the form of the differential rotation and the
meridional flows, it is unlikely in the foreseeable future that
significant constraints will be put on the transport coefficients or their
nonlinear response to the mean magnetic field.

\section{Predictions using a reconstruction of the attractor}

Having established that there are difficulties in obtaining reliable
predictions by fitting mean-field models (even if the modulation is
deterministic in origin), it is of interest to determine
whether or not more reliable predictions could be obtained by
utilising more general timeseries analysis techniques. In order to
reconstruct an attractor from a given timeseries, it is necessary to
define a corresponding phase space. There are various ways of doing
this, but given (any) discrete timeseries, $x(t)$, in which the data
is sampled at intervals of $\Delta t$, the vector 

\begin{equation} 
\mathbf{X}(t) = \left[ x(t), x(t-\Delta t)..., x(t-(d-1)\Delta t)\right]
\label{eqn:10}
\end{equation}

\noindent defines a
point in a $d$-dimensional ``embedded'' phase space (see, e.g., Farmer
\& Sidorowich 1987; Casdagli 1989). Given a time $T$, the idea of a
prediction algorithm is to find a mapping $f$ such that $f
\left(\mathbf{X}(T)\right)$ gives a good approximation to $x(T+\Delta
t)$. The predictive mapping technique that is used here uses a local
approximation method (see, e.g., Casdagli 1989), which considers the
behaviour of the nearest neighbours, in phase space, to
$\mathbf{X}(T)$. By using a least squares fit, the subsequent
evolution of each of these neighbouring points in phase space is used
to construct a piecewise-linear approximation to the predictive map,
$f$. This approximate mapping can then be applied to $\mathbf{X}(T)$
to obtain an estimate for $x(T+\Delta t)$. This algorithm can then be
repeated to find estimates for $x(T+2 \Delta t)$ and subsequent
points. The optimal value for $d$ can be determined by minimising the
error of this predictive algorithm over the known segment of the
timeseries.

The results of applying this predictor algorithm to the target
solution are also shown in Figure~5, where the timeseries
predictions are shown as crosses. The prediction is started from the
cycle maximum before the mean-field predictor diverges from the
target. Longer training timeseries lead to a more densely-populated
reconstructed attractor, which increases the probability of making
more accurate predictions. However, rather than using the
entire target run, these predictions are based upon (approximately)
$50$ cycles --- this will give a fairer comparison between these
results and timeseries predictions that are based upon the real sunspot
data. The application of the algorithm to earlier segments of
the timeseries suggests that a value of $d \ge 5$ is required in
order to minimise predictive errors. As can be seen from Figure~5,
this algorithm appears to predict the magnitude of the maximum of the
following cycle to a reasonable degree of accuracy, although the
predictions subsequently diverge from the target. Whilst neither of
these techniques are capable of producing reliable long-term predictions,
these results do suggest that for the short-term prediction of solar
magnetic activity, timeseries analysis techniques may provide a
viable alternative to predictions based simply upon mean-field dynamo
models (provided stochastic effects can be neglected). 

\section{Conclusions}

Solar magnetic activity arises as a result of a hydromagnetic dynamo
--- that much we believe  to be true. As yet, there is no consensus on
the location of the dynamo, the dominant nonlinear or stochastic
effects, or even the fundamental processes that are responsible for
the operation of such a dynamo. Although plausible mechanisms have
been proposed, as yet none of these are entirely satisfactory. Against
this background, there is a drive to be able to predict solar activity
with greater accuracy, due to the importance of this activity in
driving solar events.  

What we have demonstrated here is that no meaningful predictions can
be made from illustrative mean-field models, no matter how they are
constructed. If the mean-field model is constructed to be a driven linear
oscillator then the small stochastic effects that lead to the modulation
will have an extremely large
effect on the basic cycle and make even short-term prediction
extremely difficult. The second scenario, where the modulation arises
as a result of nonlinear processes rather than stochastic
fluctuations, is clearly a better one for prediction --- though here
too, prediction is fraught with difficulties. Owing to the inherent
nonlinearity of the dynamo system, long-term predictions are
impossible (even if the form of the model is completely correctly
determined). Furthermore, even short-term prediction from mean-field
models is meaningless because of fundamental uncertainties in the form
and amplitude of the transport coefficients and nonlinear
response. Any deterministic nonlinear model that produces chaotically
modulated activity cycles will be faced with the same difficulties. 

The equations that describe dynamo action in the solar interior are
known to be nonlinear partial differential equations --- the momentum
equation is nonlinear in both the velocity and the magnetic field. One
indication of the role played by nonlinear effects in the solar dynamo
is the presence of cyclic variations in the solar differential
rotation (the ``torsional oscillations''). Furthermore estimates of
the field strength at the base of the convection zone consistent with
the observed formation of active regions yield fields of sufficient
strength ($10^4-10^5$G) for the nonlinear Lorentz force to be
extremely significant, whilst the flows are vigorously nonlinear and
turbulent. It therefore seems extremely unlikely that the dynamics of
the solar interior can be described by a forced linear system without
throwing away much (if not all) of the important physics. In this case
it must be argued {\it not only} that this discarded physics is
irrelevant to the dynamo process but also that the parameterisation of
the unresolved physics should not include a stochastic component, as
this would have an extremely large effect on such a relinearised system.

It is certainly tempting to try to use the observed magnetic flux at
the solar surface as an input to a model for prediction (whether
nonlinear or stochastic, mean-field or full MHD). Certainly any fully
consistent solar activity model constructed in the future should be
capable of reproducing the observed pattern of magnetic activity at
the solar surface, although this will require a complete understanding
not only of the generation process via dynamo action, but also the
processes which lead to the formation and subsequent rise of
concentrated magnetic structures from the solar interior to the
surface.  However it is not clear what role the flux at the solar
surface plays in the basic dynamo process. Is it inherent to the
process (as modelled by flux transport dynamos) or simply a by-product
of the dynamo process that is occurring deep within the sun? Estimates
suggest that between $5$ and $10 \%$ of the solar flux generated in
the deep interior makes it to the solar surface (e.g. Galloway \&
Weiss 1981). For the flux at the solar surface to be the key for
dynamo action, it must be explained why the majority of the magnetic
flux that resides in the solar interior plays such a little part in
the dynamics (to such an extent that it does not even appear as a
small stochastic perturbation to the large-scale flux transport
dynamo). 

Finally it is important to stress that {\it even if} a model has been
tuned so as to reproduce results over a number of solar activity
cycles, then there is a good chance of error in the prediction for the
next cycle. Any advection-diffusion system in which one is free to
specify  not only the sources and the sinks but also the transport
processes can be tuned to reproduce any required features of activity.
Moreover, the formulation of a prediction in terms of a parameterised 
mean-field model does not inherently put the prediction on a sounder
scientific basis than a prediction based on methods of timeseries
analysis alone (some of which use very sophisticated mathematical
techniques). This, of course, is not to say that any given prediction
from such a model will be incorrect, just that the basis for making
the prediction has no strong scientific support.

\acknowledgments

We would like to thank Nigel Weiss for useful discussions and for
providing helpful comments and suggestions. PJB would like to
acknowledge the support of PPARC.   

\clearpage

\clearpage

\begin{figure}
\epsscale{0.68}
\plotone{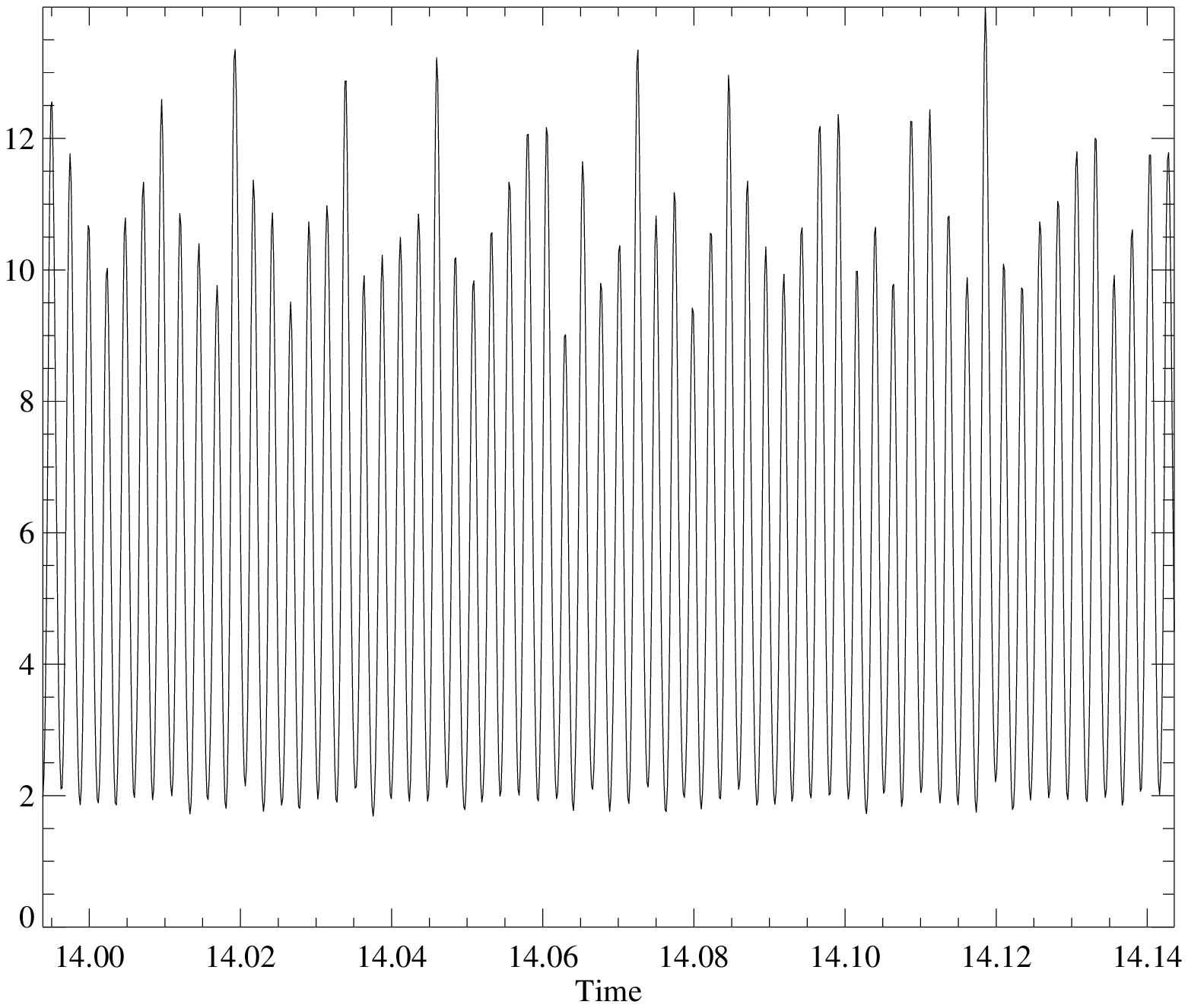}
\epsscale{0.68}
\plotone{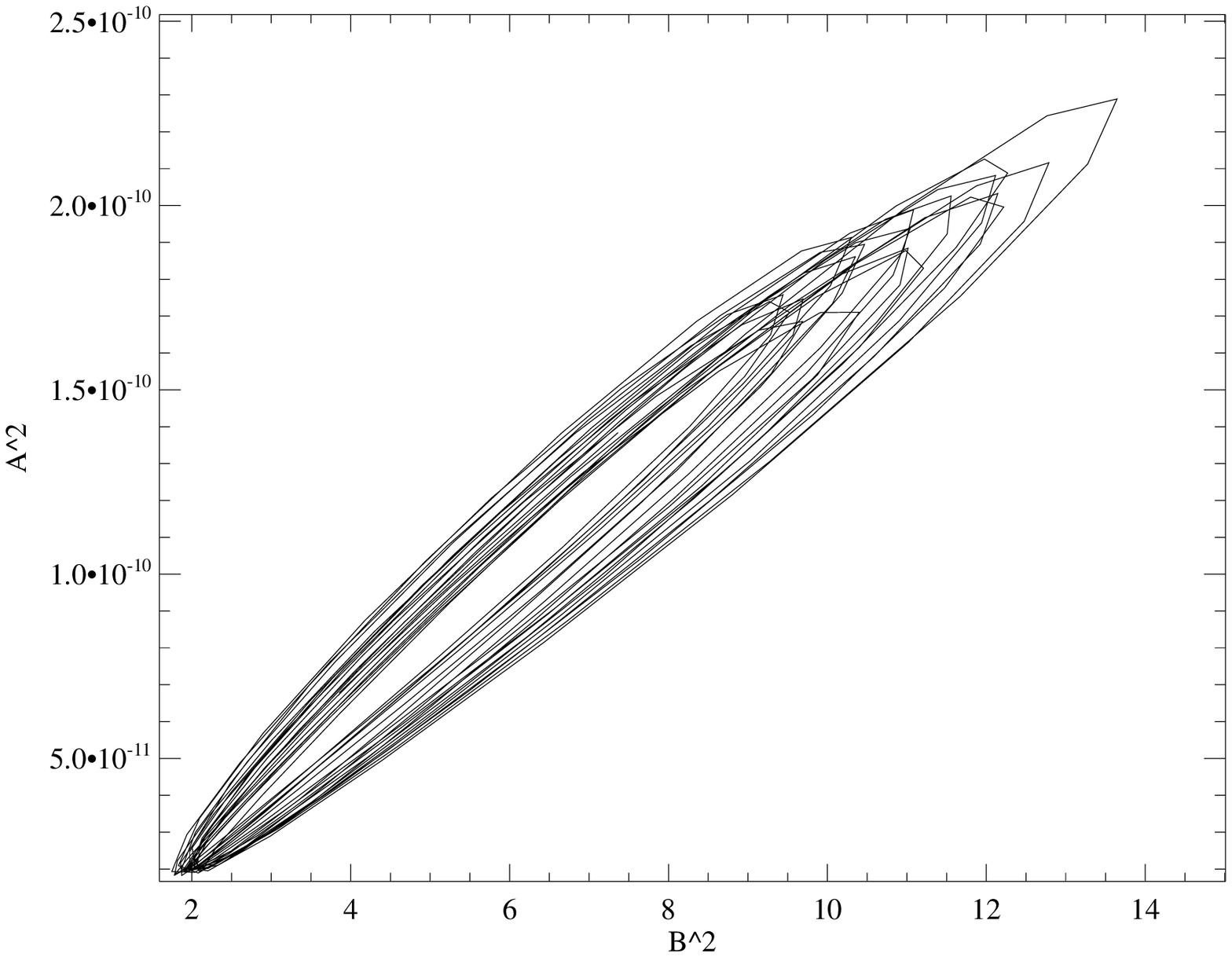}
\caption{The time evolution of the stochastically-perturbed flux
  transport dynamo. Top: Timeseries
  for the mean of the squared toroidal field ($B^2$) at the base of
  the convection zone. Bottom: In this figure, $B^2$ at the base of
  the convection zone is plotted
  against the mean of the squared values of the poloidal magnetic
  potential ($A^2$) at the surface of the domain.\label{fig1}}    
\end{figure}

\begin{figure}
\plotone{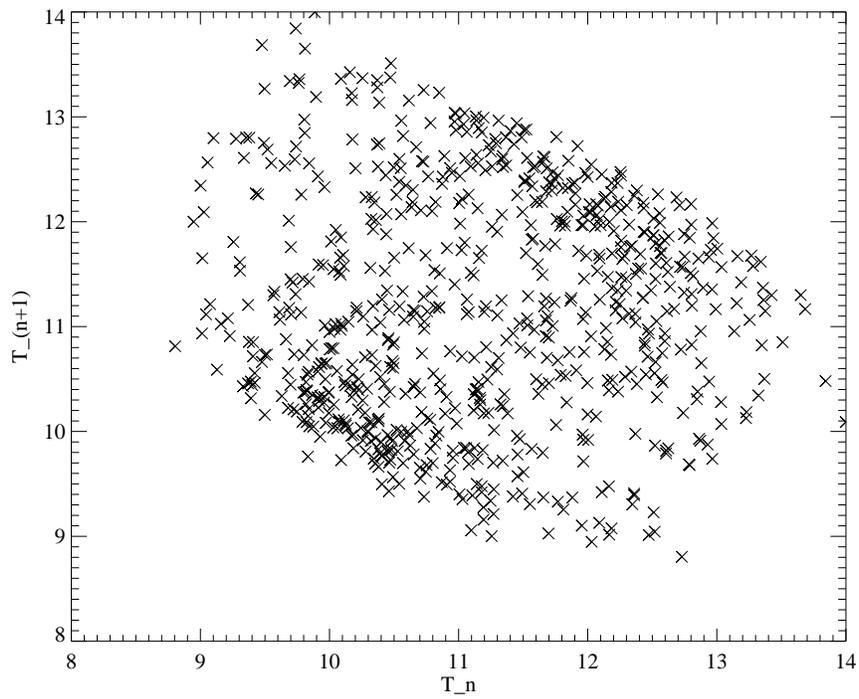}
\caption{The lack of correlation between successive maxima in the
  stochastically perturbed timeseries (as shown in Figure~1). Defining
  $T_n$ to be magnitude of the $n^{th}$ maximum, this plot shows the
  sequential behaviour of these maxima, plotting $T_{n+1}$ as a
  function of $T_n$.\label{fig2}}  
\end{figure}

\clearpage

\begin{figure}
\plotone{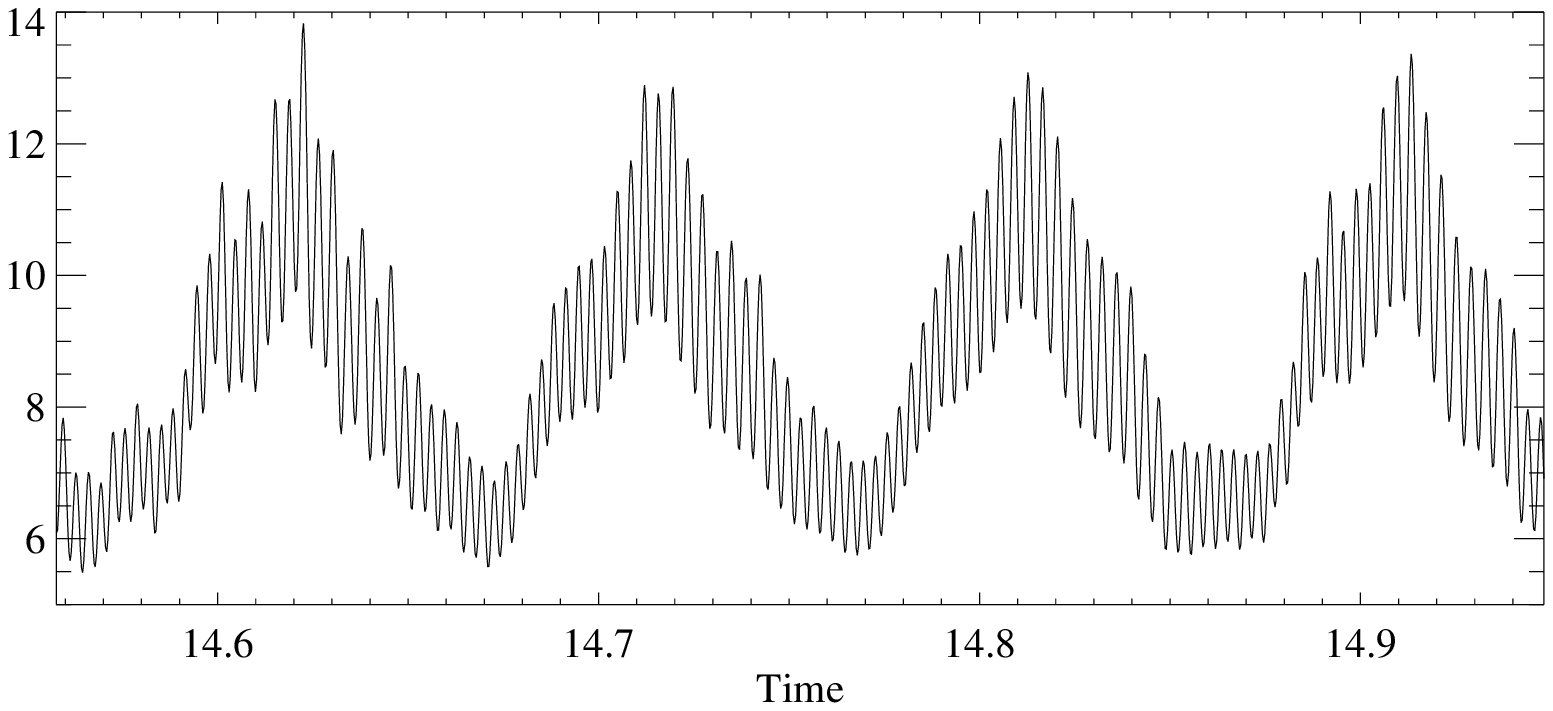}
\plotone{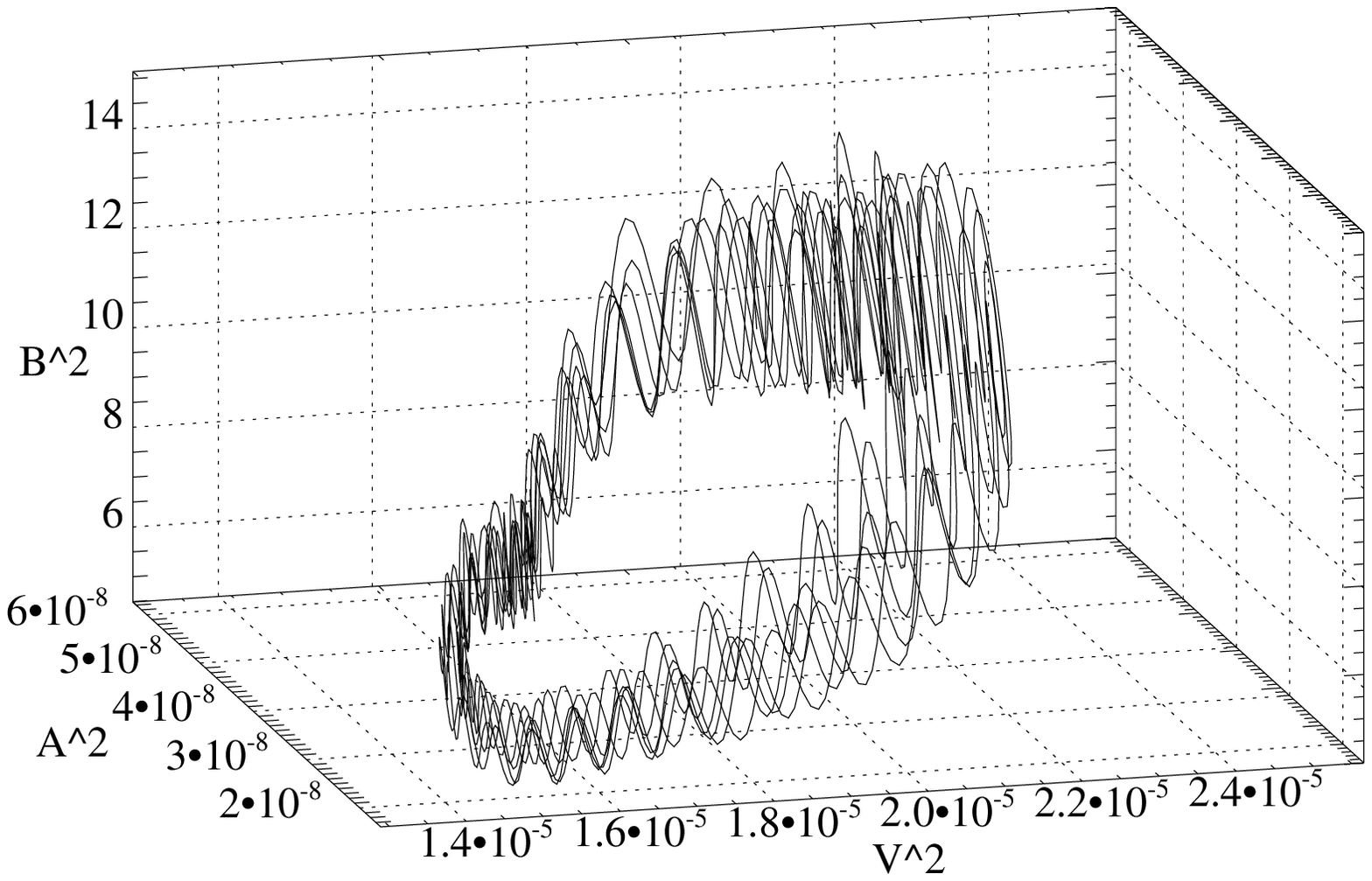}
\caption{The time evolution of the target solution. Top: Timeseries
  for the mean of the squared toroidal field ($B^2$) in the dynamo
  region. Bottom: An attractor for the target solution,
  in which $B^2$ is plotted against the mean of the squared values of the
  poloidal magnetic potential ($A^2$) and the velocity
  perturbation ($V^2$).\label{fig3}}   
\end{figure}

\clearpage

\begin{figure}
\plotone{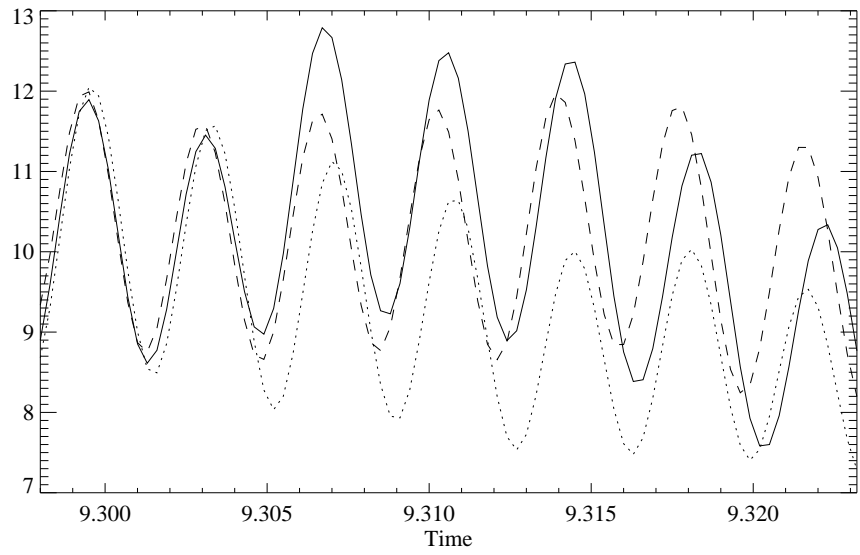}
\caption{Three timeseries showing the evolution of the mean squared
  toroidal field in the dynamo region. The solid line shows a
  segment of the target solution timeseries; the dashed and dotted
  lines show the time-evolution of solutions that are started from nearby
  points on the same attractor. Although all solutions have the same
  model parameters, the timeseries rapidly diverge after a couple of
  cycles.\label{fig4}} 
\end{figure}

\clearpage

\begin{figure}
\begin{center}
\epsscale{0.99}
\plotone{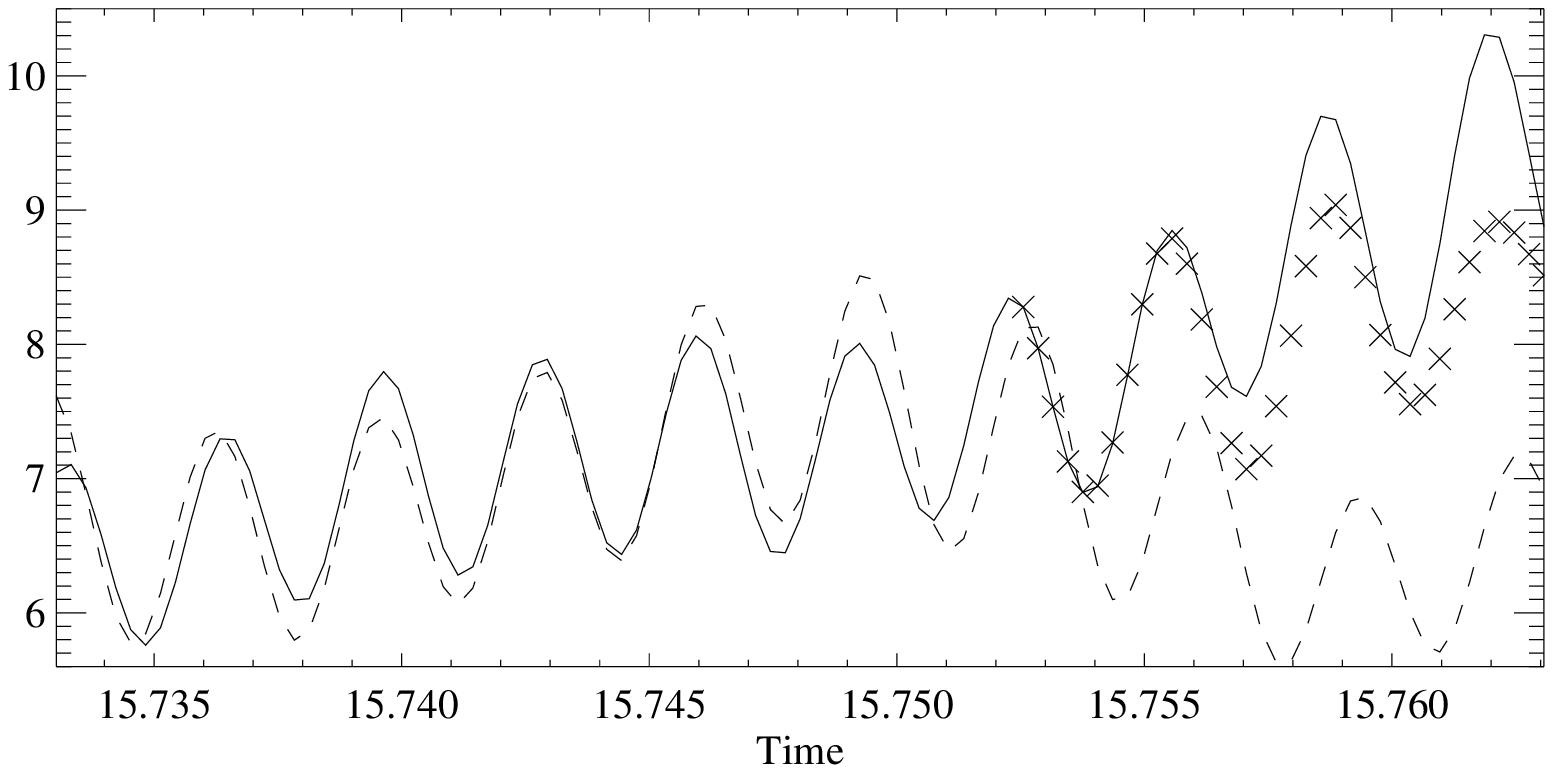}
\vspace{0.1cm}
\plotone{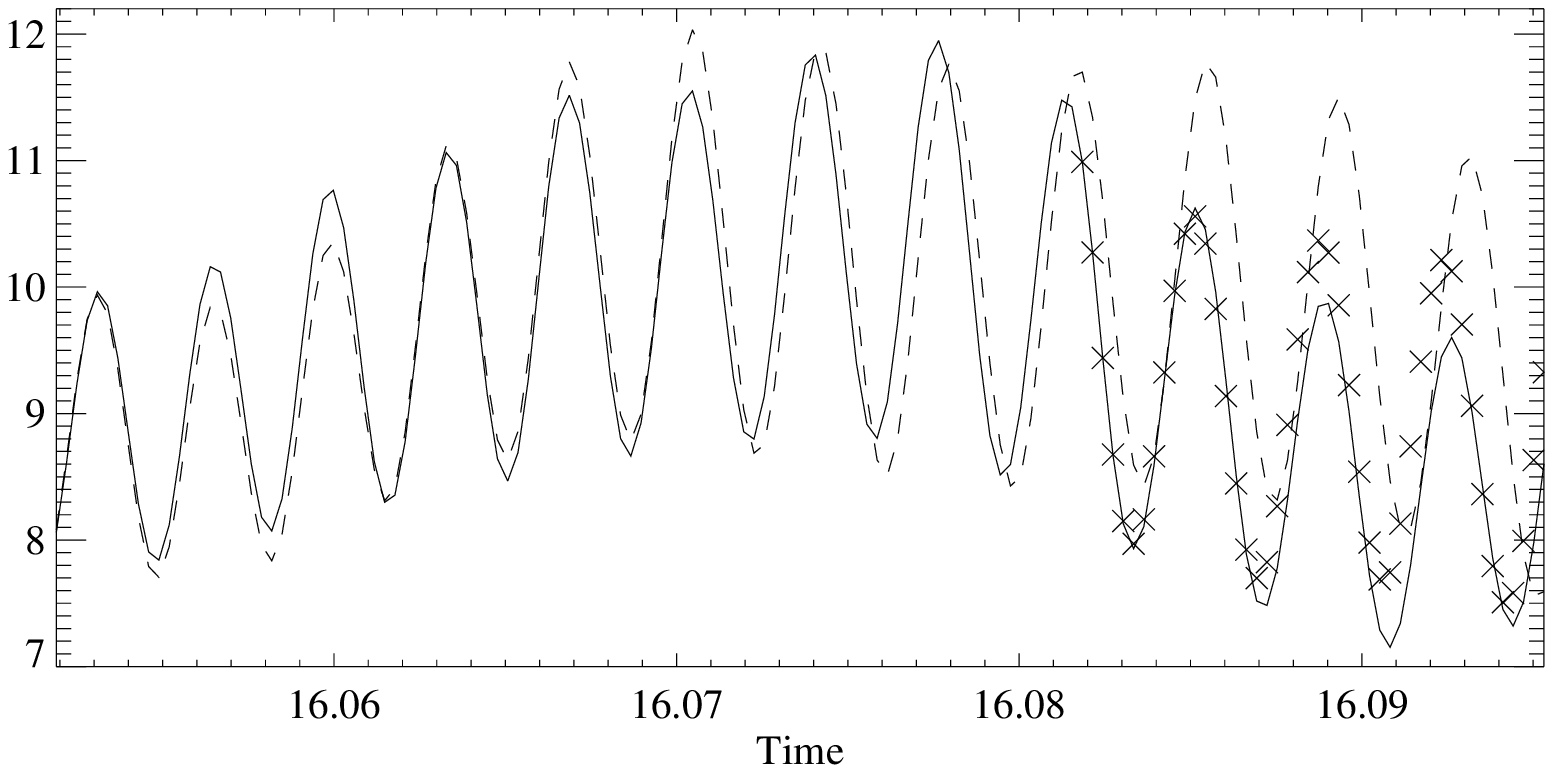}
\caption{Attempts to predict two different segments of the target
solution timeseries: In each plot, the solid line shows a segment of
the target solution timeseries, the dashed line shows the behaviour of
the chosen mean-field predictor (different predictors are used for
each plot), whilst the crosses show the predictions that are obtained
by reconstructing the nonlinear attractor for the target solution. In each
case, the mean-field predictors have been optimised by ensuring that
the chosen predictor closely matches the target timeseries segment for
a large number of cycles (6 cycles in the upper plot, 9 in the
lower).\label{fig5}}   
\end{center}
\end{figure}

\end{document}